\documentclass[aps,pra,groupedaddress,twocolumn]{revtex4}

\usepackage{amsmath,amsfonts,amssymb,graphics,graphicx,epsfig,color,times,bbm,natbib}

\begin{document}

\title{Precision of single-qubit gates based on Raman transitions}

\author{Xavier Caillet and Christoph Simon}\email{christoph.simon@ujf-grenoble.fr}
\affiliation{Laboratoire de Spectrom\'{e}trie Physique, CNRS -
Universit\'{e} de Grenoble 1, St. Martin d'H\`{e}res, France}

\date{\today}

\begin{abstract}
We analyze the achievable precision for single-qubit gates that
are based on Raman transitions between two near-degenerate ground
states via a virtually excited state. In particular, we study the
errors due to non-perfect adiabaticity and due to spontaneous
emission from the excited state. For the case of non-adabaticity,
we calculate the error as a function of the dimensionless
parameter $\chi=\Delta \tau$, where $\Delta$ is the detuning of
the Raman beams and $\tau$ is the gate time. For the case of
spontaneous emission, we give an analytical argument that the gate
errors are approximately equal to $\Lambda \gamma/\Delta$, where
$\Lambda$ is the rotation angle of the one-qubit gate and $\gamma$
is the spontaneous decay rate, and we show numerically that this
estimate holds to good approximation.
\end{abstract}


\maketitle

\section{Introduction}

The ability to perform arbitrary unitary transformations on
individual qubits is very important in the context of quantum
computation \cite{nielsenchuang}. From the point of view of
decoherence, it is often advantageous to use near-degenerate
ground states of a given system as qubits, e.g. different
hyperfine levels in a trapped ion or atom, or the spin states of
an individual excess electron in a quantum dot. In such
situations, the use of optical Raman transitions can be an
attractive approach for realizing single-qubit operations, cf.
Ref. \cite{ions} for experiments with trapped ions, Ref.
\cite{atoms} for a proposal involving atoms in an optical lattice,
and Ref. \cite{imamoglu} for a proposal with spins in quantum
dots.

A concrete procedure for realizing arbitrary single-spin
operations via Raman transitions has recently been proposed in the
context of single spins in quantum dots \cite{chen}. Variations
involving different excited states of the quantum dots (light-
hole excitons instead of heavy-hole excitons) were discussed in
\cite{calarco} and \cite{nazir}. Let us note that there are
several proposals on how to realize optically controlled two-qubit
gates between individual spins in different quantum dots
\cite{imamoglu,calarco,nazir,pier}. Schemes for qubit measurement
have also been proposed \cite{imamoglu,calarco,liu}.

In Ref. \cite{chen} it was suggested to perform the Raman
operation in an adiabatic fashion, in order to minimize the
population in the excited state, and thus gate errors due to the
decoherence of that state, which is in general much faster than
the decoherence of the near-degenerate ground states. Note that
stimulated Raman adiabatic passage (STIRAP) is commonly used in
atomic and molecular physics for the coherent transfer of quantum
states \cite{bergmann}, a task that is somewhat less general than
the realization of arbitrary single-qubit operations. For an
alternative proposal for single-qubit gates based on STIRAP see
Ref. \cite{kis}.

In a real experiment, the adiabatic approximation will never be
perfectly valid. Furthermore, spontaneous emission is an
unavoidable error mechanism in any Raman system. In quantum dots,
the interaction of the exciton with phonons can also be important,
however it is possible to fabricate dots where spontaneous
emission dominates all other sources of decoherence
\cite{dephasing}. In Ref. \cite{chen} the authors briefly
discussed the conditions for adiabaticity and the effect of
spontaneous emission from the excited state. In the present work
we perform a more detailed study of these fundamental sources of
error for the proposed gate protocol. In particular, we obtain
quantitative results for the errors due to non-adiabaticity, and a
simple formula for the errors due to spontaneous emission, namely
that they are approximately equal to $\Lambda \gamma/\Delta$,
where $\Lambda$ is the rotation angle for the single-qubit
rotation, $\gamma$ is the spontaneous emission rate and $\Delta$
is the detuning of the two Raman lasers from the excited state. We
give evidence for this result both with a simple formal argument
and by numerical computation.

This paper is organized as follows. In section \ref{protocol} we
describe the protocol for realizing arbitrary single-qubit gates
via Raman transitions proposed in Ref. \cite{chen}. In section
\ref{non-adiabatic} we study the errors due to non-perfect
adiabaticity (in the absence of spontaneous emission). In section
\ref{sp-emission} we study the errors caused by spontaneous
emission. In section \ref{discussion} we give our conclusions.

\section{Gate protocol}
\label{protocol}

\begin{figure}
\begin{center}
\includegraphics[width=0.7 \columnwidth]{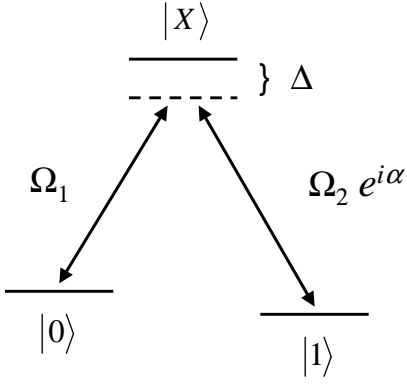}
\caption{Scheme for single qubit gates analyzed in this paper. The
low lying states $|0\rangle$ and $|1\rangle$ serve as qubit
states. They are coupled to an excited state $|X\rangle$ via two
laser beams with time-dependent Rabi frequencies $\Omega_1$ and
$\Omega_2 e^{i\alpha}$. The laser beams have the same constant
detuning $\Delta$ from the excited state. } \label{resonance}
\end{center}
\end{figure}

In this section we describe the gate protocol proposed in Ref.
\cite{chen}. Consider a three level system composed of the two
logical states of the qubit, $\left|0\right\rangle$ and
$\left|1\right\rangle$, and an auxiliary excited state
$\left|X\right\rangle$. The protocol relies on adiabatic Raman
transitions in this $\Lambda$ system, cf. Fig. \ref{resonance},
using two phase-locked laser pulses. The two laser frequencies are
chosen such that they have the same detuning $\Delta$, i.e. they
satisfy the Raman resonance condition.

The Hamiltonian in the interaction picture is:

\begin{equation}
H=\left(       \begin{array}{clrr}
     0 & 0 &  \Omega_{1}(t)e^{i\alpha}\\
     0  & 0  & \Omega_{2}(t) \\
     \Omega_{1}(t)e^{-i\alpha} & \Omega_{2}(t) & \Delta
     \end{array} \right),
\end{equation}

where $\alpha$ is the relative phase between the two real Rabi
frequencies
 $\Omega_{1}(t),\Omega_{2}(t)$.
 This Hamiltonian can be diagonalized straightforwardly. One introduces the following parameters:

\begin{equation}
\Omega(t)= \sqrt{\Omega^{2}_{1}(t)+\Omega^{2}_{2}(t)}
\end{equation}

\begin{equation}
Z(t)=\sqrt{\Omega^{2}(t)+\left( \frac{\Delta}{2}\right)^{2}      }
\end{equation}

\begin{equation}
\phi(t)=\frac{1}{2}\arctan\left(2\frac{ \Omega(t)}{\Delta}\right)
\end{equation}

\begin{equation}
\beta(t)=\arctan\left(\frac{ \Omega_{2}(t)}{\Omega_{1}(t)}\right)
\end{equation}

The angle $\beta$ is maintained constant by choosing the same
envelope shape for the two pulses. One obtains the eigenvalues :

\begin{itemize}
    \item

    $\lambda_{1}(t)=0$ with eigenvector
        \begin{equation}
     \left|\Phi_{1}(t)\right\rangle=\left(       \begin{array}{clrr}              -e^{i\alpha}\sin(\beta) \\
    \cos(\beta) \\
    0    \end{array} \right)
    \end{equation}
\end{itemize}

\begin{itemize}
    \item $\lambda_{2}(t)=-2 \:Z(t)\sin^{2}(\phi(t))$ with eigenvector
        \begin{equation}
     \left|\Phi_{2}(t)\right\rangle=\left(       \begin{array}{clrr}              -e^{i\alpha}\cos(\beta)\cos(\phi) \\
    -\sin(\beta)\cos(\phi) \\
    \sin(\phi)    \end{array} \right)
    \label{eq0}
        \end{equation}
\end{itemize}

\begin{itemize}
    \item $\lambda_{3}(t)=2 \:Z(t)\cos^{2}(\phi(t))$ with eigenvector
        \begin{equation}
     \left|\Phi_{3}(t)\right\rangle=\left(       \begin{array}{clrr}              e^{i\alpha}\cos(\beta)\sin(\phi) \\
    \sin(\beta)\sin(\phi) \\
    \cos(\phi)    \end{array} \right)
    \label{eqphi3}
        \end{equation}
\end{itemize}

The first eigenstate $\left|\Phi_{1}(t)\right\rangle$ is time
independent and completely decoupled from the other two
eigenstates. It has no contribution from the excited state
$\left|X\right\rangle$. The second eigenstate
$\left|\Phi_{2}(t)\right\rangle$ possesses only a small component
of the excited state $\left|X\right\rangle$, as long as
$\Omega/\Delta$ is small. The last state
$\left|\Phi_{3}(t)\right\rangle$ is mainly composed of the excited
state $\left|X\right\rangle$.

An arbitrary unitary transformation can be realized adiabatically
in the following way. Before the lasers are turned on (i.e. for
$\phi=0$), the initial qubit state can be expressed as a linear
combination of the first two eigenstates. By applying the two
lasers, the Hamiltonian is then changed continuously. The
adiabatic theorem states that, if the change of the Hamiltonian is
sufficiently slow, the population in each instantaneous eigenstate
remains constant, only the relative phases of the eigenstates
change. In the subspace formed by the first two eigenstates, one
obtains the following transformation:

    \begin{equation}
\left[a,b\right] \longmapsto  \left[a,b\:e^{-i\Lambda_{2}}\right]
    \end{equation}

with

    \begin{equation}
\Lambda_{2}=\int_{t_i}^{t_f}    \lambda_{2}(u)du, \label{eqAngle}
    \end{equation}
where $t_i$ and $t_f$ denote the initial and final times
respectively. From the point of view of the qubit basis spanned by
the states $|0\rangle$ and $|1\rangle$, the resulting
transformation has the form
\begin{equation}
U=e^{-i/2 \Lambda_2 \vec{\sigma} \cdot \vec{n}},
\end{equation}
where $\vec{\sigma}$ is the vector of Pauli matrices,
corresponding to a rotation through an angle $\Lambda_{2}$ about
an axis described by a unit vector $\vec{n}$ with components
    \begin{eqnarray}
n_{x}&=&\cos(\alpha)\sin(2\:\beta) \nonumber\\
n_{y}&=&-\sin(\alpha)\sin(2\:\beta)\\
n_{z}&=&\cos(2\:\beta). \nonumber
    \end{eqnarray}
We can now begin our detailed study of the corrections to this
idealized description under realistic experimental conditions. We
will start with errors due to non-perfect adiabaticity.

\section{Errors due to non-adiabaticity}
\label{non-adiabatic}

\subsection {Exact equations of motion}

The exact wave function can be expanded in terms of the previously
defined instantaneous eigenstates, but with in general
time-dependent coefficients:
    \begin{equation}
\psi(t)=\sum_{n}     a_{n} (t) \left|\Phi_{n}(t)\right\rangle\:
\exp\left[-i\int^{t}_{t_{i}}   \lambda_{n}(u)du\right]
    \end{equation}
Writing out the Schr\"{o}dinger equation for this wave function,
one obtains the following evolution equations for the
coefficients:
    \begin{eqnarray}
&&\dot{ a}_{k} \exp\left[-i\int^{t}_{t_{i}}
\lambda_{k}(u)du\right]=\\
&&-\sum_{n}  a_{n} (t)\:\left\langle \Phi_{k}(t) \| \dot{
\Phi}_{n}(t)\right\rangle\: \exp\left[-i\int^{t}_{t_{i}}
\lambda_{n}(u)du\right]\nonumber
    \end{eqnarray}
Substituting the values of the scalar products $\left\langle
\Phi_{i}(t) \| \dot{ \Phi}_{j}(t)\right\rangle $ according to the
definition of the eigenstates, one finally has:

    \begin{eqnarray}
\dot{ a}_{1}(t)&=&0 \nonumber\\ \dot{ a}_{2}(t)&=&
-\dot{\phi}(t)\:a_{3}(t)\:P_{23}(t)\\
\dot{ a}_{3}(t)&=&\dot{\phi}(t)\:a_{2}(t)\:P_{32}(t)\nonumber
    \end{eqnarray}

with

    \begin{eqnarray}
 P_{32}(t)=   P^{*}_{23}(t)  &&=  \exp\left( i\int^{t}_{t_{i}}  \left(   \lambda_{3}(v)- \lambda_{2}(v)        \right)   dv  \right)  \nonumber \\  &&= \exp\left( 2i\int^{t}_{t_{i}}  Z(v)dv
 \right).
    \end{eqnarray}
    The resolution of this
system of differential equations allows to determine the error due
to non-adiabaticity.

For a given desired transformation $U$, we will define the error
as the maximum departure (in terms of overlap) of the real final
state $\rho(t_f)$ from the ideal final state
$|\psi_{ideal}\rangle=U|\psi(t_i)\rangle$, where the maximization
is over all initial states:
\begin{equation}
E(U)=\max \limits_{\psi(t_i)} \,\, \left[1-\left\langle
\psi_{ideal}\right|\rho(t_f)
\left|\psi_{ideal}\right\rangle\right]. \label{error}
    \end{equation}
The error can be expressed in terms of the complex coefficients
$a_i(t_f)$. Since the coefficient $a_1$ always remains constant,
$a_3(t_i)=0$ and the differential equations are linear and
homogeneous, it is in fact sufficient to solve the system of
equations for one initial value of $a_2$, say $a_2(t_i)=1$. The
results for all possible initial states can then simply be
obtained by multiplying the solution with the corresponding value
of $a_2(t_i)$.

\subsection {Calculation of the error due to non-adiabaticity}

We now proceed to calculate the error due to non-perfect
adiabaticity. To simplify the discussion, we will only consider
laser pulse shapes $f(t)$ that are approximate Gaussians of
halfwidth $\tau$ centered at $t=0$, slightly modified such that
the Rabi frequency is exactly zero at the initial and final times
($t_i$ and $t_f$). We introduce the ratio
    \begin{equation}
x(t)=\frac{\Omega(t)}{\Delta}  =x_{max}f(t),
    \end{equation}
where $x_{max}$ is the maximal value of the ratio $\Omega/\Delta$,
i.e. we have normalized $f$ such that $f(0)=1$.

To gain a better comprehension of the adiabatic approximation, we
make the following substitutions : for each function $g$ of the
time $t$, we write $\tilde{g}(u)=g(\tau u)$, with the
correspondence $t=\tau u$. Thus the new system evolves between the
unitless time $u_i=t_{i}/\tau$ and $u_{f}=t_{f}/\tau$.

Introducing the dimensionless quantity
    \begin{equation}\chi=\Delta\tau,
    \end{equation}
i.e. the product of the detuning and the gate time, we obtain the
functions
    \begin{equation}
\tilde{P}_{32}(u)=\exp
\left(-i\chi\int^{u}_{u_i}\sqrt{1+4\:x^{2}_{max}\tilde{f}^{2}(v)}dv\right)
    \end{equation}
and
    \begin{equation}
\dot{\tilde{ \phi  }}(u)=
\frac{x_{max}\dot{\tilde{f}}(u)}{1+4\:x^{2}_{max}\tilde{f}^{2}(u)},
    \end{equation}
which appear in the dimensionless evolution equations:

    \begin{equation}
\dot{\tilde{ a}}_{2}(u)= \dot{\tilde{ \phi  }}(u)
\:\tilde{a}_{3}(u)\:     \tilde{P}^{*}_{32}(u)
    \end{equation}

        \begin{equation}
\dot{\tilde{ a}}_{3}(u)= -\dot{\tilde{ \phi  }}(u)
\:\tilde{a}_{2}(u)\:     \tilde{P}_{32}(u)
    \end{equation}

We thus obtain a new system of differential equations depending on
the two dimensionless parameters $ \chi$ et $x_{max}$. We are
interested in the dependence of the solutions on the two
parameters. First it appears that the greater $\chi$ is, the
faster the term $\tilde{P}_{32}$ is oscillating, and thus the less
the population $\left|a_{3}\right|^{2}$ is important. We can also
obtain a reduction of $\left|a_{3}\right|^{2}$ by reducing
$\dot{\tilde{ \phi  }}$. This can be done by decreasing $x_{max}$.
This preliminary analysis suggests that the error decreases with
$\chi$ and increases with $x_{max}$.

\begin{figure}
\begin{center}
\includegraphics[width=0.9 \columnwidth]{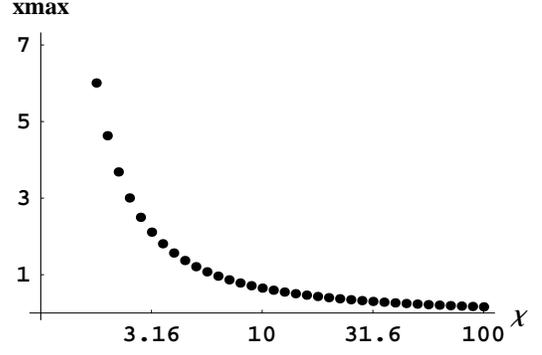}
\caption{Determination of $x_{max}$, the maximum value of the
ratio $\Omega/\Delta$, as a function of the dimensionless
parameter $\chi=\Delta \tau$ for a $\pi$ rotation. The values for
other rotation angles can be obtained from the fact that $x_{max}$
depends only on the ratio $\chi/\Lambda$.} \label{figx}
\end{center}
\end{figure}

In the following, we will study the gate error as a function of
the rotation angle $\Lambda$ and the dimensionless parameter
$\chi$. The quantity $x_{max}$ is then not an independent
variable, but is determined by these two parameters in the
following way. After simplifications and substitutions, Eq.
(\ref{eqAngle}) becomes :
 \begin{equation}
2    \Lambda  =  \chi  \int^{u_f }_{u_i }\left(
\sqrt{1+4\:x^{2}_{max}\tilde{f}^{2}(u)}  -1      \right)  du =\chi
\: g(x_{max}). \label{lambda}
    \end{equation}
Recall that $\tilde{f}(u)$ is essentially a Gaussian with
halfwidth 1 (apart from a small modification at the boundaries of
the time interval), normalized such that its maximum value is
equal to one. This equation gives $x_{max}$ as an implicit
function of the ratio $\Lambda/\chi$. As $g$ is an increasing
function, and is a bijection from $\left[0,+\infty\right]$ to
$\left[0,+\infty\right]$, we obtain a one to one correspondence
between $x_{max}$ and $\chi$ for a given $\Lambda$. Fig.
\ref{figx} shows that $x_{max}$ is a decreasing function of
$\chi$.

 \begin{figure}
\begin{center}
\includegraphics[width=0.9 \columnwidth]{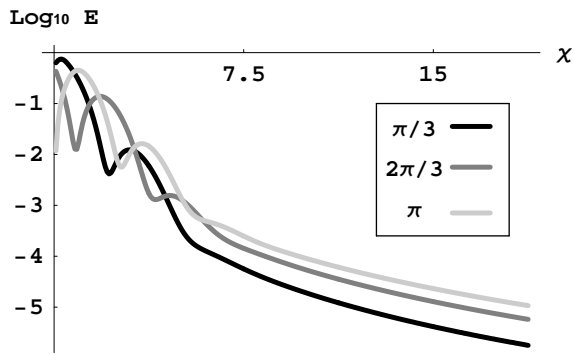}
\caption{Gate error as a function of $\chi$ for different rotation
angles.} \label{figDiffRot}
\end{center}
\end{figure}

For a given rotation angle $\Lambda$, we finally obtain a system
depending only on the parameter $\chi$. By solving it numerically,
we obtain an estimation of the adiabatic error as a function of
$\chi$ and of $\Lambda$. Note that the error does not depend on
the axis of rotation. Indeed, the choice of the axis of rotation
only determines the relation of the basis of logical states to the
basis of adiabatic eigenstates, and since the error is obtained by
maximizing over all initial states, this choice has no effect on
its value. The error as a function of $\chi$ for different values
of the angle $\Lambda$ is shown in Fig. \ref{figDiffRot}. For each
value of $\Lambda$, we observe two characteristic regimes. The
first one, where $\chi$ is small, is characterized by a damped
oscillatory behaviour as a function of $\chi$, leading also to a
non-monotonous variation of the error with the rotation angle. In
the second one, the error decreases continuously with $\chi$, and
greater values of $\Lambda$ lead to larger errors.

In order to minimize the error due to non-adiabaticity,
$\chi=\Delta\tau$ should thus be as large as possible. However, it
should be kept in mind that $\tau$ corresponds to the time of the
gate and thus has to be much shorter than the decoherence time of
the qubit states, in order to limit errors due to decoherence. In
principle one can be in the adiabatic regime even for very short
gate times, provided that the detuning $\Delta$ is made large
enough. However, this requires an increase in the laser amplitude
$\Omega$, in order to still achieve the same rotation angle
$\Lambda$. This relationship is contained in Eq. (\ref{lambda})
and Fig. \ref{figx}, which show that for fixed $\chi$ the rotation
angle is determined by $x_{max}$, i.e. the ratio of $\Omega$ and
$\Delta$ at maximum laser intensity. If one decreases $\tau$ and
increases $\Delta$, keeping $\chi$ and thus the level of error
constant, one therefore has to increase $\Omega$ by the same
factor as $\Delta$, in order to keep the rotation angle constant.
Since the laser intensity cannot be made arbitrarily large, this
imposes an upper bound on $\Delta$, and thus a lower bound on
$\tau$. From Fig. \ref{figDiffRot} one can see that for
$\chi\geq15$, the error is significantly less than $10^{-4}$,
which should be small enough for fault tolerant quantum
computation \cite{nielsenchuang}.

\section{Errors due to spontaneous emission}
\label{sp-emission}

\subsection {Estimate of error based on population transfer}
\label{estimate}

In this section we will investigate the error introduced to the
Raman single-qubit gates by the finite lifetime of the excited
state $\left|X\right\rangle$ due to spontaneous emission. We will
begin with a fairly simple argument that gives the correct
behaviour for the error, before presenting more precise numerical
calculations in the next subsection.

As described before, the two adiabatic basis states that are
significantly populated during the gate operation are
$\left|\Phi_{1}(t)\right\rangle $ and
$\left|\Phi_{2}(t)\right\rangle$. The state
$\left|\Phi_{1}(t)\right\rangle $ has no contribution from the
excited state and is thus unaffected by spontaneous emission. On
the other hand, the state $\left|\Phi_{2}(t)\right\rangle $ has a
component in the excited state $\left|X\right\rangle$. From Eq.
(\ref{eq0}), we  can see that the population in the unstable state
$\left|X\right\rangle$ is therefore $ a^{2}_{2}(t)
\sin^{2}(\phi(t)) $. Defining $\gamma $ as the spontaneous decay
rate of the state $\left|X\right\rangle$, we can then estimate the
population $\delta$ transferred by spontaneous emission during the
gate operation as follows:

    \begin{equation}
\delta=\int^{t_{f}}_{t_{i}} \gamma \: a^{2}_{2}(t)
\sin^{2}(\phi(t)) dt \label{eqdelta}
    \end{equation}

Let us assume that we are in the adiabatic regime ($\chi\geq 15$),
and that the overall error due to spontaneous emission is small.
The first of these conditions implies that $x_{max}$ and thus
$\phi$ is small, cf. Fig. \ref{figx}. The second one implies that
$a_{2}$ is nearly constant. Eq. (\ref{eqdelta}) can then be
simplified to
    \begin{equation}
\delta=  \gamma \: \tau \: a^{2}_{2}\: x^{2}_{max}
\int^{u_{f}}_{u_{i}}
 \tilde{f}^{2}(u)  du, \label{eqdeltasimple}
    \end{equation}
where we have again introduced the dimensionless function
$\tilde{f}(u)$ defined above. Furthermore, in the same regime, Eq.
(\ref{lambda}) can be simplified to
    \begin{equation}
x^{2}_{max} \int^{u_{f} }_{u_{i} } \tilde{f}^{2}(u)du
=\frac{\Lambda}{\chi}.
    \label{eq100}
    \end{equation}
Choosing $a^{2}_{2}=1$ in order to obtain an upper bound, we find
the following expression for the total transferred population due
to spontaneous emission:
    \begin{equation}
\delta= \Lambda \frac{\gamma}{\Delta} \label{delta}
    \end{equation}
The error induced by spontaneous emission is twofold; on the one
hand, a new distribution of the populations
$\left|a^{2}_{i}\right|$, and on the other hand, a dephasing
between the two qubit basis states. The transferred population
$\delta$ provides an estimate for the gate error due to
spontaneous emission. It may seem surprising that $\delta$ does
not depend on the gate time $\tau$, even though for longer gate
times the component of the system in the excited state has more
time to decay. The reason for this is that for the same rotation
angle shorter gate times require larger populations in the excited
state, and the two effects cancel out exactly, at least within the
framework of the above estimate. We are now going to use numerical
computation to obtain more precise results on the gate errors.

\subsection {Master equation}

In order to study the effects of spontaneous emission on the Raman
gate protocol in detail, we use the master equation formalism. In
the present section, we assume for simplicity that the spontaneous
emission can only occur from the state $\left|X\right\rangle$
toward the two qubit states $\left|0\right\rangle$ and
$\left|1\right\rangle$, and not to any additional states. To take
this decay into account, we introduce the Lindblad operators :

    \begin{equation}
 L_{i}=\sqrt{\gamma_{i}} \sigma_{i}
    \end{equation}

with $\sigma_{i}=\left|i\right\rangle \left\langle X\right| $ (for
$i=0,1$). The constants $\gamma_{i}$ are the decay rates from
$|X\rangle$ towards the states $|i\rangle$. The total decay rate
is thus given by $\gamma=\gamma_{1}+\gamma_{2}$.

The master equation \cite{nielsenchuang} is:
   \begin{equation}
\frac{d\rho}{dt} = -i\left[H,\rho\right]  +   \sum_{i}  \left(
2L_{i}\rho L^{+}_{i}-\left\{L^{+}_{i}L_{i},\rho \right\}\right).
    \end{equation}
For quantum gate operations, the initial state
$\left|\psi\right\rangle$ is always a linear combination of the
two logical states $a\left|0\right\rangle+b\left|1\right\rangle$
with $\left|a\right|^{2}+\left|b\right|^{2}=1$.

The master equation corresponds to a set of coupled differential
equations for the elements of the density matrix that can be
solved numerically. This allows us to determine the gate error
defined in Eq. (\ref{error}).

\subsection {Example - Populations and Purity}

In this subsection, we describe the numerical results for a
particular case in detail, comparing the situations with and
without spontaneous emission. This is intended to serve as an
introduction and illustration for our more general results
presented in the next subsection.

We consider a rotation by $\pi$ along the $x$ axis, corresponding
to $\alpha=0$ and $\beta= \pi /4$. . The initial state of the
system is the state $\left|0\right\rangle$. The ideal final state
is the state $\left|1\right\rangle$. We choose the values
$\tau=0.01$ ns, $\Delta=1$ meV$\approx1500 \:\mbox{ns}^{-1}$ and
$\gamma_{1}=\gamma_{2}=20\:ns^{-1}$. While these values could
apply to conceivable experiments with quantum dots
\cite{decayrate}, we have also chosen them such that the relevant
effects are clearly visible. For these values $\chi=15$. The
adiabatic approximation is thus very well satisfied, cf. Fig.
\ref{figDiffRot}.

 \begin{figure}
\begin{center}
\includegraphics[width=0.9 \columnwidth]{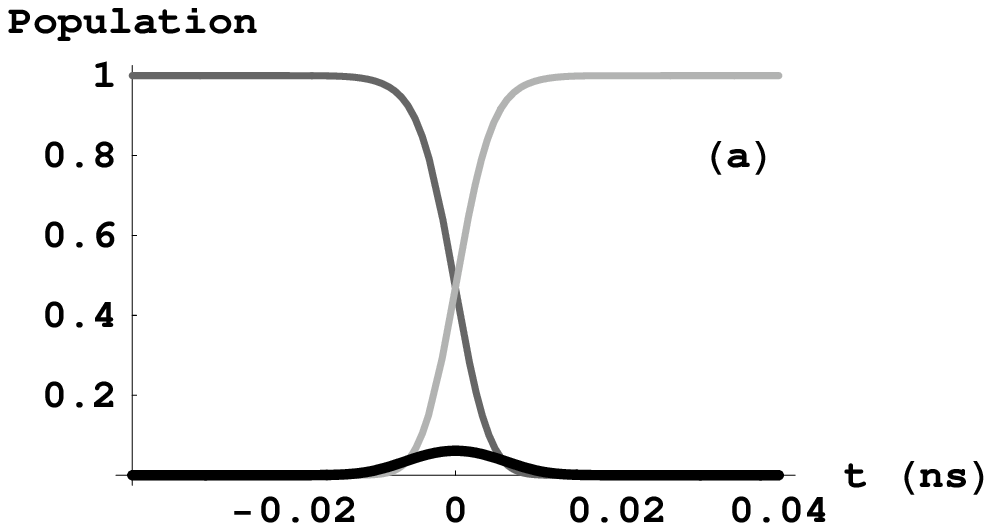}
\includegraphics[width=0.9 \columnwidth]{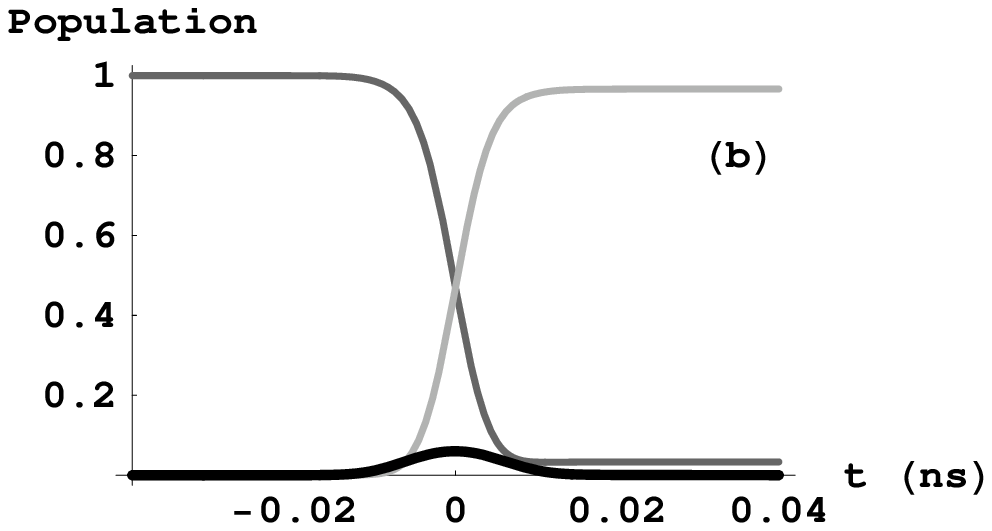}
\caption{Evolution of the populations during a $\pi$ rotation of
the state $|0\rangle$ around the $x$ axis into the state
$|1\rangle$: (a) without spontaneous emission; (b) with
spontaneous emission.} \label{populations}
\end{center}
\end{figure}

Fig. \ref{populations}(a) shows the time evolution of the
populations in the states $|0\rangle$, $|1\rangle$ and $|X\rangle$
without spontaneous emission. As expected, the populations in the
states $|0\rangle$ and $|1\rangle$ are exchanged. Moreover, the
population in $|X\rangle$ is zero at the end of the operation. One
can describe the evolution of the system in detail as follows. The
initial state $\left|0\right\rangle$ can be expressed in the
adiabatic basis as:
\begin{equation}
 \left|0\right\rangle=-\frac{1}{\sqrt{2}}\left(
\left|\Phi_{1}(t_{i})\right\rangle +
\left|\Phi_{2}(t_{i})\right\rangle \right)
\end{equation}
The final state $\left|1\right\rangle$ can be written :
\begin{equation}
 \left|1\right\rangle=\frac{1}{\sqrt{2}}\left(
\left|\Phi_{1}(t_{f})\right\rangle -
\left|\Phi_{2}(t_{f})\right\rangle \right)
\end{equation}
with
$\left|\Phi_{1}(t_{f})\right\rangle=\left|\Phi_{1}(t_{i})\right\rangle$
and
$\left|\Phi_{2}(t_{f})\right\rangle=\left|\Phi_{2}(t_{i})\right\rangle$.
The transformation is adiabatic : expressed in the basis ($
\left|\Phi_{1}(t)\right\rangle$,$
\left|\Phi_{2}(t)\right\rangle$), the state of the system is :
\begin{equation}
 \left|\Phi(t)\right\rangle =-\frac{1}{\sqrt{2}}[1,e^{i \Lambda
(t)}].
\end{equation}
While the lasers are on, the phase $\Lambda (t)$ grows and thus
the state $ \left|0\right\rangle$ is continuously transformed into
the state $ \left|1\right\rangle$. The state
$\left|\Phi_{2}(t)\right\rangle$ has a component of order
$x_{max}\:f(t)$ in the state $\left|X\right\rangle$, the
population in the excited state therefore grows until the maximum
of the light intensity is reached, and then returns to zero. Fig.
\ref{populations}(b) shows the same populations in the presence of
spontaneous emission.  It appears clearly that the rotation is no
longer perfect in this case. The populations in the initial state
$|0\rangle$ and in the excited state $|X\rangle$ are no longer
zero at the moment when the light is turned off. Of course the
remaining population in $|X\rangle$ will decay towards the states
$|0\rangle$ and $|1\rangle$ on the larger timescale set by
$\gamma_{1,2}$.

\begin{figure}
\begin{center}
\includegraphics[width=0.9 \columnwidth]{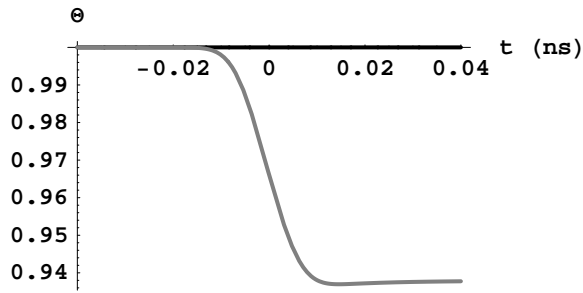}
\caption{Evolution of the purity of the state, $\Theta(t)$,
without and with spontaneous emission.} \label{theta}
\end{center}
\end{figure}

The departure from the perfect rotation can be further visualized
by analyzing the purity of the system density matrix $\rho (t)$,
i.e. by studying the quantity $\Theta(t)=\mbox{Tr} \rho(t)^2$.
Fig. \ref{theta} shows the evolution of $\Theta$ with and without
spontaneous emission. As expected, $\Theta$ remains constant in
the absence of spontaneous emission. In the presence of
spontaneous emission, the purity decreases considerably. This
decrease is particularly strong around $t=0$, i.e. around the
maximum of the laser intensity. It is at this point that the
population in the state $|X\rangle$ becomes the largest. The
spontaneous emission from $|X\rangle$ causes the state to become
mixed.

 \begin{figure}
\begin{center}
\includegraphics[width=0.9 \columnwidth]{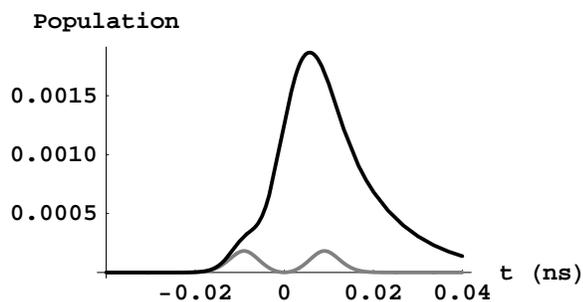}
\caption{Time evolution of the population in the adiabatic basis
state $\left|\Phi_{3}(t)\right\rangle$ (whose dominant component
is in the excited state $|X\rangle$) without and with spontaneous
emission.} \label{phi3}
\end{center}
\end{figure}

Fig. \ref{phi3} shows the time evolution of the population in the
state $ \left|\Phi_{3}(t)\right\rangle$ of Eq. (\ref{eqphi3}). One
sees that in the absence of spontaneous emission the adiabatic
approximation is well justified, the population in the state
remains very small. Its departure from zero corresponds to the
error due to non-perfect adiabaticity discussed in the previous
section. On the other hand, the presence of spontaneous emission
causes transitions between the states
$\left|\Phi_{2}(t)\right\rangle$ and
$\left|\Phi_{3}(t)\right\rangle$, which populate the latter. This
population decreases on the timescale of the radiative lifetime,
since $\left|\Phi_{3}(t)\right\rangle$ contains predominantly the
excited state $|X\rangle$.

\subsection {General results on errors due to spontaneous emission}

\begin{figure}
\begin{center}
\includegraphics[width=0.9 \columnwidth]{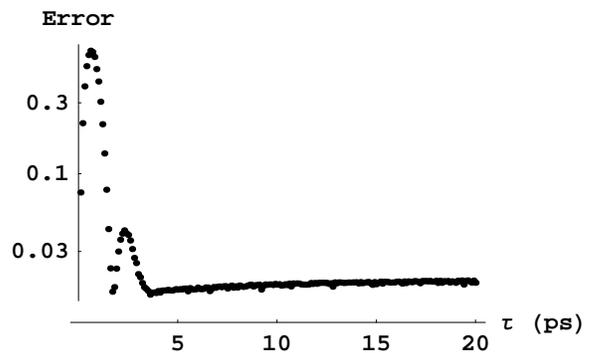}
\caption{Gate error in the presence of spontaneous emission as a
function of the gate time $\tau$. The spontaneous decay rates are
$\gamma_{1}=\gamma_{2}=5\:$ns$^{-1}$, the detuning
$\Delta=1\:$meV. The variation of $\tau$ between 0 and 20 ps
corresponds to the dimensionless parameter $\chi=\Delta \tau$
varying from 0 to 30. } \label{errorwithchi}
\end{center}
\end{figure}

In this subsection we will present more general results on the
error due to spontaneous emission. In particular we want to test
the estimate made in subsection \ref{estimate}. Fig.
\ref{errorwithchi} shows the error for a rotation by $\pi$ as a
function of the gate time $\tau$ in the presence of a spontaneous
emission. This graph should be compared to Fig. \ref{figDiffRot},
which shows the same quantity in the absence of spontaneous
emission. For short times the behaviour is very similar, showing
the same damped oscillatory character. In this regime, the error
is dominated by non-perfect adiabaticity. For longer gate times,
there is a clear difference. In the presence of spontaneous
emission, the error does not fall below a certain minimal value
and is virtually independent of $\tau$. This is in good
correspondence with the prediction made in subsection
\ref{estimate}.

 \begin{figure}
\begin{center}
\includegraphics[width=0.9 \columnwidth]{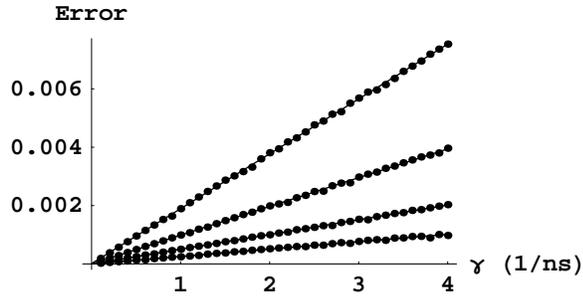}
\caption{Gate error for a $\pi$ rotation as a function of the
spontaneous decay rate $\gamma$ for fixed values of the detuning
$\Delta=1,2,4,8$ meV respectively (from top to bottom graph).}
\label{gamma}
\end{center}
\end{figure}

\begin{figure}
\begin{center}
\includegraphics[width=0.9 \columnwidth]{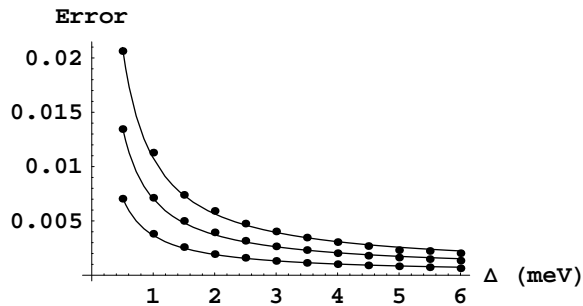}
\caption{Gate error of a $\pi$ rotation as a function of detuning
$\Delta$ for fixed values of the spontaneous decay rate
$\gamma=2,4,6$ ns$^{-1}$ respectively (from bottom to top graph).
The curves are fits to a $1/\Delta$ behaviour.} \label{Delta}
\end{center}
\end{figure}

\begin{figure}
\begin{center}
\includegraphics[width=0.9 \columnwidth]{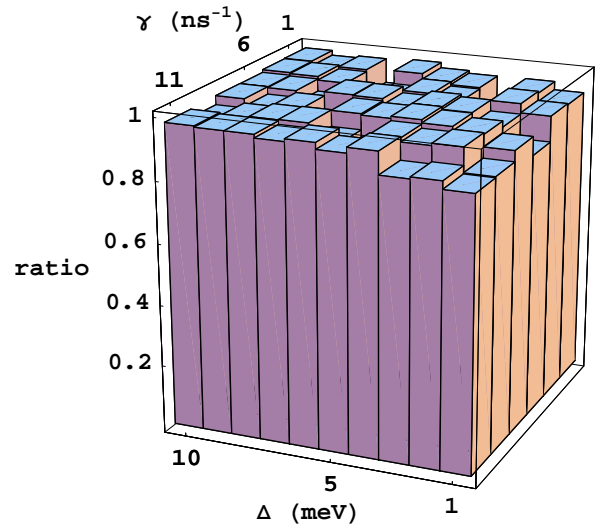}
\caption{Ratio of the (numerically obtained) exact value of the
error for a $\pi$ rotation to the value of $\pi \gamma/\Delta$
given by the estimate of subsection \ref{estimate} for a range of
values of $\gamma$ and $\Delta$.} \label{pref}
\end{center}
\end{figure}

In order to test the above estimate more systematically, we have
performed calculations varying $\gamma$ and $\Delta$ for a fixed
value of the gate time $\tau=13.3$ ps. We restrict ourselves to
values of $\Delta \geq 1$ meV, corresponding to $\chi \geq 20$, in
order to make sure that the error due to non-adiabaticity is
negligible, cf. Fig. \ref{figDiffRot}. Furthermore we focus on the
regime where the overall gate error is at most at the percent
level, since this is the relevant regime for quantum computing.
The total spontaneous decay rate is $\gamma=\gamma_1+\gamma_2$.
For simplicity we have again chosen $\gamma_1=\gamma_2$.

Fig. \ref{gamma} shows the behaviour of the gate error of a $\pi$
rotation as a function of the spontaneous decay rate $\gamma$ for
four different values of the detuning $\Delta$. One sees that the
error is linear in $\gamma$ with a high degree of accuracy.
Similarly, Fig. \ref{Delta} shows the gate error as a function of
$\Delta$ for three different values of $\gamma$. One sees that the
results are fitted extremely well by a $1/\Delta$ behaviour. The
proportionality of the error to $\gamma/\Delta$ is thus seen to be
very well obeyed. To assess the accuracy of the above estimate
concerning the absolute size of the error, Fig. \ref{pref}
compares the errors obtained numerically for a $\pi$ rotation to
the estimated error of $\pi \gamma/\Delta$. One sees that the
approximation works very well in the considered regime. As
expected, it tends to work somewhat less well for increasing
values of $\gamma$ and decreasing values of $\Delta$, i.e.
increasing overall size of the error, cf. subsection
\ref{estimate}.

\section{Conclusions}
\label{discussion}

The results obtained in the present paper give quantitative
information for the implementation of quantum computing using the
considered gate protocol. The results of section
\ref{non-adiabatic} on the errors due to non-perfect adiabaticity
make it possible to determine the maximum allowable gate speed for
any desired level of error. Our analysis also shows that it is in
principle possible to perform gates in the adiabatic regime even
for very short gate times by increasing the detuning $\Delta$.
However, this requires a corresponding increase in laser power.

The results of section \ref{sp-emission} quantify the errors due
to spontaneous emission, whose presence is unavoidable in any gate
scheme based on Raman transitions. In the context of quantum
computing with spins in quantum dots, the present analysis
complements the results of Refs. \cite{calarco,roszak} on quantum
gate errors due to phonon-induced dephasing. Phonon-related errors
can be made essentially arbitrarily small by making the gate
operation slower. The basic reason for this is that the speed of
the operation determines the energy that is available for the
creation of phonons (since the system does not decay from the
excited state during the dephasing). The slower the operation, the
less energy is available, restricting the available state space
for phonon creation. Unfortunately there is no corresponding
energy constraint for spontaneous emission, since the energy for
photon creation is provided by the decay of the emitter to one of
the low-lying states. Our results show explicitly that slowing
down the gate operation is not helpful to reduce errors in the
present context, see Fig. \ref{errorwithchi}.

On the other hand, our analysis also shows that it is possible to
choose long gate times even in the presence of spontaneous
emission, without significantly changing the size of the error due
to the decay. Relatively long gate times can be advantageous
because they allow greater frequency selectivity in schemes based
on spectral addressing of different qubits. Of course, the gate
time always has to be much shorter than the decoherence time of
superpositions of the qubit states.

\acknowledgements

We are grateful to J. Eymery, J.-M. G\'{e}rard, Y.-M. Niquet and
J.-P. Poizat for useful discussions.

\bibliographystyle{apsrev}

\end{document}